\documentclass[pra,showpacs,superscriptaddress]{revtex4}
\usepackage{amssymb,amsmath,amsfonts}
\usepackage{times}
\usepackage{bbm,bm}
\usepackage{graphicx}
\newcommand{\p}{{\bm r}}

\newcommand{\tr}{\mathrm{tr}}
\newcommand{\ra}{\rangle}

\newcommand{\su}{{\bm u}}
\newcommand{\sv}{{\bm v}}
\newcommand{\jv}{{\bm j}}
\newcommand{\iv}{{\bm i}}
\newcommand{\kv}{{\bm k}}
\newcommand{\lm}{\Lambda_{\max}^2}

\begin{document}

\title{Geometric measure of entanglement and shared quantum states}

\author{Levon Tamaryan}

\affiliation{Physics Department, Yerevan State University,
Yerevan, 375025, Armenia}


\author{DaeKil Park}

\affiliation{Department of Physics, Kyungnam University, Masan,
631-701, Korea}


\author{Jin-Woo Son}

\affiliation{Department of Mathematics, Kyungnam University,
Masan, 631-701, Korea}


\author{Sayatnova Tamaryan}

\affiliation{Theory Department, Yerevan Physics Institute,
Yerevan, 375036, Armenia}



\begin{abstract}
We give an explicit expression for the geometric measure of
entanglement for three qubit states that are linear combinations
of four orthogonal product states. It turns out that the geometric
measure for these states has three different expressions depending
on the range of definition in parameter space. Each expression of
the measure has its own geometrically meaningful interpretation.
Such an interpretation allows oneself to take one step toward a
complete understanding for the general properties of the
entanglement measure. The states that lie on joint surfaces
separating different ranges of definition, designated as shared
states, seem to have particularly interesting features. The
properties of the shared states are fully discussed.
\end{abstract}

\maketitle

\section{Introduction}

Entanglement is the most intriguing feature of quantum mechanics
and a key resource in quantum information science. One of the main
goals in these theories is to develop a comprehensive theory of
multipartite entanglement. Various entanglement measures have been
invented to quantify the multi-particle entanglement
\cite{dist,form,vedr,relative,eis,wall,wei} but none of them were able to
suggest a method for calculating a measure of
multipartite systems. This mathematical difficulty is the main
obstacle to elaborate a theory of multi-particle entanglement.

In this paper, we present the first calculation of the geometric
measure of entanglement \cite{wei,Shim,barn} for three qubit
states which are expressed as linear combinations of four given
orthogonal product states. Any pure three qubit state can be
written in terms of five preassigned orthogonal product states
\cite{acin} via Schmidt decomposition. Thus the states discussed
here are more general states compared to the well-known GHZ
\cite{ghz} and W \cite{w} states.

The reason for using the geometric measure of entanglement is that
it is suitable for any partite system regardless of its
dimensions. However, analytical computation for generic states
still remains as a great challenge. The measure depends on
entanglement eigenvalue $\lm$ and can be derived from the formula
$E_g(\psi)=1-\lm$. For pure states, the entanglement eigenvalue is
equal to the maximal overlap of a given state with any complete
product state. This measure has the following remarkable
properties:

\smallskip

i) it has an operational treatment. The same overlap $\lm$ defines
Groverian measure of entanglement \cite{bno,mix-bno} which has been introduced
later in operational terms. In other words, it quantifies how well a given state serves
as an input state to Grover's search algorithm \cite{grov}. From this
view, Groverian measure can be regarded as an operational treatment of the
geometric measure.

\smallskip

ii) it has identified irregularity in
channel capacity additivity \cite{wern}. Using this measure, one
can show that a family of quantities, which were thought to be
additive in an earlier papers, actually are not. For
example, it is natural to conjecture that preparing two pairs of
entangled particles should give us twice the entanglement of one
pair and, similarly, using a channel twice doubles its capacity.
However,  this conjecture claiming additivity has proved to
be wrong in some cases.

\smallskip

iii) it has useful connections to other entanglement measures and
gives rise to a lower bound on the relative entropy of
entanglement \cite{con1} and generalized robustness \cite{con2}.
For certain pure states the first lower bound is saturated and
thus their relative entropy of entanglement can be deduced
from their geometric measure of
entanglement. The second lower bound to generalized robustness can
be express in terms of $\lm$ directly.
\smallskip

Owing to these features, the geometric measure can play an
important role in the investigation of different problems related
to entanglement. For example, the entanglement of two distinct
multipartite bound entangled states can be determined analytically
in terms of a geometric measure of entanglement \cite{bound}.
Recently, the same measure has been used to understand the
physical implication of Zamolodchikov's c-theorem \cite{zamo86}
more deeply. It is an important application regarding the quantum
information techniques in the effect of renormalization group in
field theories \cite{orus07}. Thus it is natural that geometric
measure of entanglement is an object of intense interest and in
some recent works revised \cite{cao} and generalized \cite{kobes}
versions of the geometric measure were presented.

The progress made to date allows oneself to calculate the
geometric measure of entanglement for pure three qubit systems
\cite{theor}. The basic idea is to use $(n-1)$-qubit mixed states
to calculate the geometric measure of $n$-qubit pure states. In
the case of three qubits this idea converts the task effectively
into the maximization of the two-qubit mixed state over product
states and yields linear eigenvalue equations \cite{lev}. The
solution of these linear eigenvalue equations reduces to the root
finding for algebraic equations of degree six. However,
three-qubit states containing symmetries allow complete analytical
solutions and explicit expressions as the symmetry reduces the
equations of degree six to the quadratic equations. Analytic
expressions derived in this way are unique and the presented
effective method can be applied for extended quantum systems. Our
aim is to derive analytic expressions for a wider class of three
qubit systems and in this sense this work is the continuation of
Ref.\cite{lev}.

We consider most general W-type three qubit states that allow to
derive analytic expressions for entanglement eigenvalue. These
states can be expressed as linear combinations of four given
orthogonal product states. If any of coefficients in this
expansion vanishes, then one obtains the states analyzed in
\cite{lev}. Notice that arbitrary linear combinations of five product
states \cite{acin} give a couple of algebraic equations of degree
six. Hence \'Evariste Galois's theorem does not allow to get
analytic expressions for these states except some particular
cases.

We derive analytic expressions for an entanglement eigenvalue. Each
expression has its own applicable domain depending on state
parameters and these applicable domains are split up by separating
surfaces. Thus the geometric measure distinguishes different types
of states depending on the corresponding applicable domain. States
that lie on separating surfaces are shared by two types of states
and acquire new features.

In Section II we derive stationarity equations and their solutions.
In Section III we specify three qubit states under consideration
and find relevant quantities. In Section IV we calculate
entanglement eigenvalues and present explicit expressions. In
Section V we separate the validity domains of the derived
expressions. In Section VI we discuss shared states. In section
VII we make concluding remarks.

\section{Stationarity equations}

In this section we briefly review the derivation of the
stationarity equations  and their general solutions \cite{lev}.
Denote by $\rho^{ABC}$ the density matrix of the three-qubit pure
state and define the entanglement eigenvalue $\lm$ \cite{wei}

\begin{equation}\label{gen.pmax}
\lm=\max_{\varrho^1\varrho^2\varrho^3}
\tr\left(\rho^{ABC}\varrho^1\otimes\varrho^2\otimes\varrho^3\right),
\end{equation}

\noindent where the maximization runs over all normalized complete
product states. Theorem 1 of Ref.\cite{theor} states that the
maximization of a pure state over a single qubit state can be
completely derived by using  a
particle traced over density matrix. Hence the theorem allows us to
re-express the entanglement eigenvalue by reduced density matrix
$\rho^{AB}$ of qubits A and B

\begin{equation}\label{gen.pred}
\lm=\max_{\varrho^1\varrho^2}
\tr\left(\rho^{AB}\varrho^1\otimes\varrho^2\right).
\end{equation}

Now we introduce four Bloch vectors:

\smallskip

1)\,$\p_A$ for the reduced density matrix $\rho^A$ of the qubit A,

2)\,$\p_B$ for the reduced density matrix $\rho^B$ of the qubit B,

3)\,$\su$ for the single qubit state $\varrho^1$,

4)\,$\sv$ for the single qubit state $\varrho^2$.

\smallskip

Then the expression for entanglement eigenvalue (\ref{gen.pred}) takes
the form

\begin{equation}\label{gen.s1s2}
\lm=\frac{1}{4}\max_{u^2=v^2=1}\left(1+\su\cdot \p_A+\sv\cdot
\p_B+g_{ij}\,u_iv_j\right),
\end{equation}

\noindent where(summation on repeated indices $i$ and $j$ is
understood)

\begin{equation}\label{gen.vec}
g_{ij}=\tr(\rho^{AB}\sigma_i\otimes\sigma_j)
\end{equation}

\noindent and $\sigma_i$'s are Pauli matrices. The closest product
state satisfies the stationarity conditions

\begin{equation}\label{gen.eq}
\p_A+g\sv=\lambda_1\su,\quad\p_B+g^T\su=\lambda_2\sv,
\end{equation}

\noindent where Lagrange multipliers $\lambda_1$ and $\lambda_2$
enforce the unit Bloch vectors $\su$ and $\sv$. The solutions of Eq.(\ref{gen.eq}) are

\begin{equation}\label{gen.sol}
\su=\left(\lambda_1\lambda_2\openone-g\,g^T\right)^{-1}
\left(\lambda_2\p_A+g\,\p_B\right),\quad
\sv=\left(\lambda_1\lambda_2\openone-g^Tg\right)^{-1}
\left(\lambda_1\p_B+g^T\p_A\right).
\end{equation}

\noindent Unknown Lagrange multipliers are defined by equations

\begin{equation}\label{gen.alg}
u^2=1,\quad v^2=1.
\end{equation}

In general, Eq.(\ref{gen.alg}) gives algebraic equations of degree
six. The reason for this is that stationarity equations define all extremes
of the reduced density matrix $\rho^{AB}$ over product states,
regardless of them being global or local. And the degree of the algebraic
equations is the number of possible extremes.

Eq.(\ref{gen.sol}) contains valuable information. It provides solid bases
for a new numerical approach. This can be compared with the numerical calculations
based on other technique \cite{Shim-grov}.

\section{Three Qubit State}

We consider W-type state

\begin{equation}\label{w.psi}
|\psi\ra=a|100\ra+b|010\ra+c|001\ra+d|111\ra,
\end{equation}

\noindent where free parameters $a,b,c,d$ satisfy the
normalization condition $a^2+b^2+c^2+d^2=1$. Without loss of
generality we consider only the case of positive parameters
$a,b,c,d$. At first sight, it is not obvious whether the state
allows analytic solutions or not. However, it does and our first
task is to confirm the existence of the analytic solutions.

In fact, entanglement of the state Eq.(\ref{w.psi}) is invariant
under the permutations of four parameters $a,b,c,d$. The
invariance under the permutations of three parameters $a,b,c$ is
the consequence of the invariance under the permutations of qubits
A,B,C. Now we make a local unitary(LU) transformation that
relabels the bases of qubits B and C, i.e. $0_B\leftrightarrow1_B,\;
0_C\leftrightarrow1_C$, and does not change the basis of qubit A.
This LU-transformation interchanges the coefficients as follows:
$a\leftrightarrow d,\; b\leftrightarrow c$. Since any entanglement
measure must be invariant under LU-transformations and the
permutation $b\leftrightarrow c$, it must be also invariant under
the permutation $a\leftrightarrow d$. In view of this symmetry, any
entanglement measure must be invariant under the permutations of
all the state parameters $a,b,c,d$. Owing to this symmetry, the
state allows to derive analytic expressions for the entanglement
eigenvalues. The necessary condition is \cite{lev}

\begin{equation}\label{w.det}
\det\left(\lambda_1\lambda_2\openone-g\,g^T\right)=0.
\end{equation}

Indeed, if the condition (\ref{w.det}) is fulfilled, then the
expressions (\ref{gen.sol}) for the general solutions are not
applicable and Eq.(\ref{gen.eq}) admits further simplification.

Denote by $\iv,\jv,\kv$ unit vectors along axes $x,y,z$
respectively. Straightforward calculation yields

\begin{equation}\label{w.matr}
 \p_A=r_1\,\kv,\quad\p_B=r_2\,\kv,\quad g=
\begin{pmatrix}
2\omega & 0 & 0\\
0 & 2\mu & 0\\
0 & 0 & -r_3
\end{pmatrix}
,
\end{equation}
where
\begin{eqnarray}\label{w.eig}
 & & r_1=b^2+c^2-a^2-d^2,\quad r_2=a^2+c^2-b^2-d^2,\quad
r_3=a^2+b^2-c^2-d^2,\\\nonumber
 & & \hspace{3.0cm} \omega=ab+dc,\quad\mu=ab-dc.
\end{eqnarray}

Vectors $\su$ and $\sv$ can be written as linear combinations

\begin{equation}\label{w.ijk}
\su=u_i\iv+u_j\jv+u_k\kv,\quad \sv=v_i\iv+v_j\jv+v_k\kv
\end{equation}

\noindent of vectors $\iv,\jv,\kv$. The substitution of the
Eq.(\ref{w.ijk}) into Eq.(\ref{gen.eq}) gives a couple of
equations in each direction. The result is a system of six linear
equations

\begin{subequations}\label{w.sub}
\begin{equation}\label{w.sub1}
2\omega\,v_i=\lambda_1u_i,\quad2\omega\,u_i=\lambda_2v_i,
\end{equation}
\begin{equation}\label{w.sub2}
2\mu\,v_j=\lambda_1u_j,\quad2\mu\,u_j=\lambda_2v_j,
\end{equation}
\begin{equation}\label{w.sub3}
r_1-r_3 v_k=\lambda_1u_k,\quad r_2-r_3 u_k=\lambda_2v_k.
\end{equation}
\end{subequations}

Above equations impose two conditions

\begin{subequations}\label{w.imp}
\begin{equation}\label{w.impom}
(\lambda_1\lambda_2-4\omega^2)u_iv_i=0,
\end{equation}
\begin{equation}\label{w.impmu}
(\lambda_1\lambda_2-4\mu^2)u_jv_j=0.
\end{equation}
\end{subequations}

From these equations it can be deduced that the condition (\ref{w.det})
is valid and the system of equations (\ref{gen.eq}) and
(\ref{gen.alg}) is solvable. Note that as a consequences of
Eq.(\ref{w.sub}) $x$ and/or $y$ components of vectors $\su$ and
$\sv$ vanish simultaneously. Hence, conditions (\ref{w.imp}) are
satisfied in following three cases:

\begin{itemize}

\item vectors $\su$ and $\sv$ lie in $xz$ plane
\begin{equation}\label{w.vers1}
\lambda_1\lambda_2-4\omega^2=0,\quad u_jv_j=0,
\end{equation}

\item vectors $\su$ and $\sv$ lie in $yz$ plane
\begin{equation}\label{w.vers2}
\lambda_1\lambda_2-4\mu^2=0,\quad u_iv_i=0,
\end{equation}

\item vectors $\su$ and $\sv$ are aligned with axis $z$
\begin{equation}\label{w.vers3}
u_iv_i=u_jv_j=0.
\end{equation}

\end{itemize}

These cases are examined individually in next section.

\section{Explicit expressions}

In this section we analyze all three cases and derive explicit
expressions for entanglement eigenvalue. Each expression has its
own range of definition in which they are deemed applicable. Three ranges of
definition cover the four dimensional sphere given by
normalization condition. It is necessary to separate the validity
domains and to make clear which of expressions should be applied for
a given state. It turns out that the separation of domains requires
solving inequalities that contain polynomials of degree six. This
is a nontrivial task and we investigate it in the next section.

\subsection{Circumradius of Convex Quadrangle}

Let us consider the first case. Our main task is to find Lagrange
multipliers $\lambda_1$ and $\lambda_2$. From equations
(\ref{w.sub3}) and (\ref{w.vers1}) we have

\begin{equation}\label{cq.sz}
u_k=\frac{\lambda_2r_1-r_2r_3}{4\omega^2-r_3^2},\quad
v_k=\frac{\lambda_1r_2-r_1r_3}{4\omega^2-r_3^2}.
\end{equation}

In its turn Eq.(\ref{w.sub1}) gives

\begin{equation}\label{cq.sx}
\lambda_1u_i^2=\lambda_2v_i^2.
\end{equation}

Eq.(\ref{gen.alg}) allows the substitution of expressions (\ref{cq.sz})
into Eq.(\ref{cq.sx}). Then we can obtain the second equation for Lagrange
multipliers

\begin{equation}\label{cq.con}
\lambda_1\left(4\omega^2+r_2^2-r_3^2 \right)=
\lambda_2\left(4\omega^2+r_1^2-r_3^2 \right).
\end{equation}

This equation has a simple form owing to condition (\ref{w.det}).
Thus we can factorize the equation of degree six into the quadratic
equations. Equations (\ref{cq.con}) and (\ref{w.vers1}) together
yield

\begin{equation}\label{cq.lam-om}
\lambda_1=2\omega\,\frac{bc+ad}{ac+bd},\quad
\lambda_2=2\omega\,\frac{ac+bd}{bc+ad}.
\end{equation}

Note that we kept only positive values of Lagrange multipliers and
omitted negative values to get the maximal value of $\lm$. Now
Eq.(\ref{gen.s1s2}) takes the form

\begin{equation}\label{cq.eigprim1}
4\lm=1+\frac{8(ab+cd)(ac+bd)(ad+bc)-r_1r_2r_3}{4\omega^2-r_3^2}.
\end{equation}

In fact, entanglement eigenvalue is the sum of two equal terms and
this statement follows from the identity

\begin{equation}\label{cq.iden}
1-\frac{r_1r_2r_3}{4\omega^2-r_3^2}=
8\frac{(ab+cd)(ac+bd)(ad+bc)}{4\omega^2-r_3^2}.
\end{equation}

\noindent To derive this identity one has to use the normalization
condition $a^2+b^2+c^2+d^2 = 1$. The identity
allows to rewrite Eq.(\ref{cq.eigprim1}) as follows

\begin{equation}\label{cq.eigfin}
\lm=4R_q^2,
\end{equation}

\noindent where

\begin{equation}\label{cq.circ}
R_q^2=\frac{(ab+cd)(ac+bd)(ad+bc)}{4\omega^2-r_3^2}.
\end{equation}

Above formula has a geometric interpretation and now we
demonstrate it. Let us define a quantity $p \equiv (a+b+c+d)/2$. Then
the denominator can be rewritten as

\begin{equation}\label{cq.her}
4\omega^2-r_3^2=16(p-a)(p-b)(p-c)(p-d).
\end{equation}

Five independent parameters are necessary to construct a convex
quadrangle. However, four independent parameters are necessary to
construct a convex quadrangle that has circumradius. For such
quadrangles the area $S_q$ is given exactly by Eq.(\ref{cq.her})
up to numerical factor, that is $S^2_q=(p-a)(p-b)(p-c)(p-d)$.
Hence Eq.(\ref{cq.circ}) can be rewritten as

\begin{equation}\label{cq.radius}
R_q^2=\frac{(ab+cd)(ac+bd)(ad+bc)}{16S^2_q}.
\end{equation}

\noindent Thus $R_q$ can be interpreted as a  circumradius of the convex
quadrangle. Eq.(\ref{cq.radius}) is the generalization of the
corresponding formula of Ref.\cite{lev} and reduces to the circumradius
of the triangle if one of parameters is zero.

Eq.(\ref{cq.eigfin}) is valid if vectors $\su$ and $\sv$ are unit
and have non-vanishing $x$ components. These conditions have short
formulations

\begin{equation}\label{cq.ineq}
|u_k|\leq1,\quad|v_k|\leq1.
\end{equation}

Above inequalities are polynomials of degree six and algebraic
solutions are unlikely. However, it is still possible do define
the domain of validity of Eq.(\ref{cq.radius}).

\subsection{Circumradius of Crossed-Quadrangle}

Here, we consider the second case given by
Eq.(\ref{w.vers2}). Derivations repeat steps of the previous
subsection and the only difference is the interchange
$\omega\leftrightarrow\mu$. Therefore we skip some obvious steps
and present only main results. Components of vectors $\su$ and
$\sv$ along axis $z$ are

\begin{equation}\label{cr.sz}
u_k=\frac{\lambda_2r_1-r_2r_3}{4\mu^2-r_3^2},\quad
v_k=\frac{\lambda_1r_2-r_1r_3}{4\mu^2-r_3^2}.
\end{equation}

The second equation for Lagrange multipliers

\begin{equation}\label{cr.con}
\lambda_1\left(4\mu^2+r_2^2-r_3^2 \right)=
\lambda_2\left(4\mu^2+r_1^2-r_3^2 \right)
\end{equation}

together with Eq.(\ref{w.vers2}) yields

\begin{equation}\label{cr.lam-mu}
\lambda_1=\pm2\mu\,\frac{bc-ad}{ac-bd},\quad
\lambda_2=\pm2\mu\,\frac{ac-bd}{bc-ad}.
\end{equation}

\noindent Using these expressions, one can derive the following
expression for entanglement eigenvalue

\begin{equation}\label{cr.eigprim2}
4\lm=
1+\frac{\lambda_2(4\mu^2+r_1^2-r_3^2)-r_1r_2r_3}{4\mu^2-r_3^2}.
\end{equation}

Now the restrictions $1/4<\lm\leq1$ derived in Ref.\cite{theor}
uniquely define the signs in Eq.(\ref{cr.lam-mu}). Right signs
enforce strictly positive fraction in right hand side of
Eq.(\ref{cr.eigprim2}). To make a right choice, we replace $d$ by
$-d$ in the identity (\ref{cq.iden}) and rewrite
Eq.(\ref{cr.eigprim2}) as follows

\begin{equation}\label{cr.eigprim3}
4\lm=\frac{1}{2}\,\frac{(ac-bd)(bc-ad)(ab-cd)}
{p(p-c-d)(p-b-d)(p-a-d)}\pm
\frac{1}{2}\,\frac{(ac-bd)(bc-ad)(ab-cd)}
{p(p-c-d)(p-b-d)(p-a-d)}.
\end{equation}


Lower sign yields zero and is wrong. It shows that reduced density
matrix $\rho^{AB}$ still has zero eigenvalue.

Upper sign may yield a true answer. Entanglement eigenvalue is

\begin{equation}\label{cr.eigfin}
\lm=4R_\times^2,
\end{equation}

\noindent where

\begin{equation}\label{cr.radius}
R_\times^2=\frac{(ac-bd)(bc-ad)(ab-cd)}{16S^2_\times},
\end{equation}

\noindent and $S^2_\times=p(p-c-d)(p-b-d)(p-a-d)$. The formula
(\ref{cr.radius}) may seem suspicious because it is not clear whether
right hand side is positive and lies in required region. To
clarify the situation we present a geometrical treatment of
Eq.(\ref{cr.radius}).

The geometrical figure $ABCD$ in Fig.1A is not a quadrangle and is
not a polygon at all. The reason is that it has crossed sides $AD$
and $BC$. We call figure $ABCD$ crossed-quadrangle in a figurative
sense as it has four sides and a cross point. Another
justification of this term is that we will compare figure $ABCD$
in Fig.1A with a convex quadrangle $ABCD$ containing the same
sides.

\begin{figure}
\includegraphics[width=10cm]{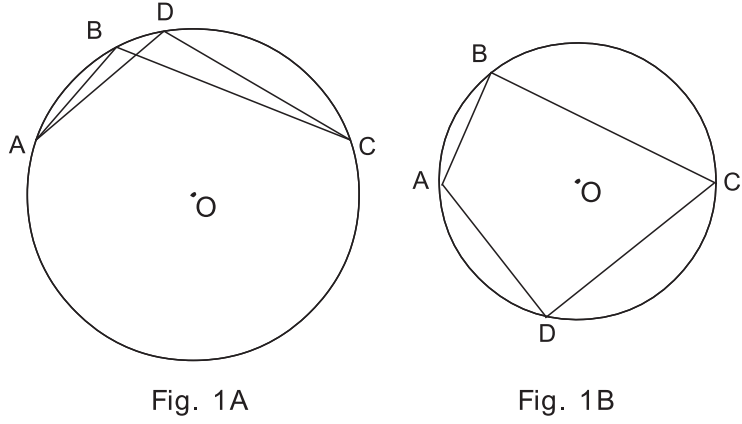}
\caption{\label{cr.eps}This figure shows the example for the case
when crossed quadrangle(Fig.1A) has larger circumradius than that
of convex quadrangle(Fig.1B) with same sides.}
\end{figure}

Consider a crossed-quadrangle $ABCD$ with sides
$AB=a,BC=b,CD=c,DA=d$ that has circumcircle. It is easy to find
the length of the interval $AC$

\begin{equation}\label{cr.ab}
AC^2=\frac{(ac-bd)(bc-ad)}{ab-cd}.
\end{equation}

This relation is true unless triangles $ABC$ and $ADC$ have the
same height and as a consequence equal areas. Note that $S_\times$
is not an area of the crossed-quadrangle. It is the difference
between the areas of the noted triangles.

Using Eq.(\ref{cr.ab}), one can derive exactly Eq.(\ref{cr.radius})
for the circumradius of the crossed-quadrangle.

Eq.(\ref{cr.eigfin}) is meaningful if vectors $\su$ and $\sv$ are
unit and have nonzero components along the axis $y$.

\subsection{Largest Coefficient}

In this subsection we consider the last case described by
Eq.(\ref{w.vers3}). Entanglement eigenvalue takes maximal value if
all terms in r.h.s. of Eq.(\ref{gen.s1s2}) are positive. Then
equations (\ref{w.vers3}) and (\ref{w.matr}) together impose

\begin{equation}\label{lc.vec}
\su={\rm Sign}(r_1)\kv,\quad\sv={\rm Sign}(r_2)\kv,\quad
r_1r_2r_3<0,
\end{equation}

\noindent where Sign(x) gives -1, 0 or 1 depending on whether x is
negative, zero, or positive. Substituting these values into
Eq.(\ref{gen.s1s2}), we obtain

\begin{equation}\label{lc.3}
\lm=\frac{1}{4}\left(1+|r_1|+|r_2|+|r_3|\right).
\end{equation}

Owing to inequality, $r_1r_2r_3<0$, above expression always gives a square of
the largest coefficient $l$

\begin{equation}\label{lc.l}
l=\max(a,b,c,d)
\end{equation}

\noindent in Eq.(\ref{w.psi}). Indeed, let us consider the case
$r_1>0,r_2>0,r_3<0$. From inequalities $r_1>0,r_2>0$ it follows
that $c^2>d^2+|a^2-b^2|$ and therefore $c^2>d^2$. Note, $c^2>d^2$
is necessary but not sufficient condition. Now if $d>b$, then
$r_1>0$ yields $c>a$ and if $d<b$, then $r_3<0$ yields $c>a$. Thus
inequality $c>a$ is true in all cases. Similarly $c>b$ and $c$ is
the largest coefficient. On the other hand $\lm=c^2$ and
Eq.(\ref{lc.3}) really gives the largest coefficient in this case.

Similarly, cases $r_1>0,r_2<0,r_3>0$ and $r_1<0,r_2>0,r_3>0$ yield
$\lm=b^2$ and $\lm=a^2$, respectively. And again entanglement
eigenvalue takes the value of the largest coefficient.

The last possibility $r_1<0,r_2<0,r_3<0$ can be analyzed using
analogous speculations. One obtains $\lm=d^2$ and $d$ is the
largest coefficient.

Combining all cases mentioned earlier, we rewrite Eq.(\ref{lc.3}) as follows

\begin{equation}\label{lc.larg}
\lm=l^2.
\end{equation}

This expression is valid if both vectors $\su$ and $\sv$ are
collinear with the axes $z$.

\bigskip

We have derived three expressions for
(\ref{cq.eigfin}),(\ref{cr.eigfin}) and (\ref{lc.larg}) for
entanglement eigenvalue. They are valid when vectors $\su$ and
$\sv$ lie in $xz$ plane, lie in $yz$ plane and are collinear with
axis $z$, respectively. The following section goes on to specify these domains by
parameters $a,b,c,d$.

\section{Applicable Domains}

Mainly, two points are being analyzed. First, we probe into the meaningful
geometrical interpretations of quantities $R_q$ and $R_\times$.
Second, we separate validity domains of equations
(\ref{cq.eigfin}),(\ref{cr.eigfin}) and (\ref{lc.larg}). It is
mentioned earlier that algebraic methods for solving the
inequalities of degree six are ineffective. Hence, we use
geometric tools that are elegant and concise in this case.

We consider four parameters $a,b,c,d$ as free parameters as the
normalization condition is irrelevant here. Indeed, one can use
the state $|\psi\ra/\sqrt{a^2+b^2+c^2+d^2}$ where all parameters
are free. If one repeats the same steps, the only difference is
that the entanglement eigenvalue $\lm$ is replaced by
$\lm/(a^2+b^2+c^2+d^2)$. In other words, normalization condition
re-scales the quadrangle, convex or crossed, so that the
circumradius always lies in the required region. Consequently, in
constructing quadrangles we can neglect the normalization condition
and consider four free parameters $a,b,c,d$.

\subsection{Existence of circumcircle.}

It is known that four sides $a,b,c,d$ of the convex
quadrangle must obey the inequality $p-l>0$. Any set of such
parameters forms a cyclic quadrilateral. Note that the quadrangle is
not unique as the sides can be arranged in different orders. But
all these quadrangles have the same circumcircle and the
circumradius is unique.

The sides of a crossed-quadrangle must obey the same condition.
Indeed, from Fig.1A it follows that $BC-AB<AC<AD+DC$ and
$DC-AD<AC<AB+BC$. Therefore $AB+AD+DC>BC$ and $AB+BC+AD>DC$. The
sides $BC$ and $DC$ are two largest sides and consequently
$p-l>0$. However, the existence of the circumcircle requires an
additional condition and it is explained here. The relation
$r_3=2\mu\cos ABC$ forces $4\mu^2\geq r_3^2$ and, therefore

\begin{equation}\label{ad.posden}
S_\times^2\geq0.
\end{equation}

\noindent Thus the denominator in Eq.(\ref{cr.radius}) must be
positive. On the other hand the inequality $AC^2\geq0$ forces a
positive numerator of the same fraction

\begin{equation}\label{ad.posnum}
(ac-bd)(bc-ad)(ab-cd)\geq0.
\end{equation}

These two inequalities impose conditions on parameters $a,b,c,d$.
For the future considerations, we need to write explicitly the
condition imposed by inequality (\ref{ad.posnum}). The numerator
is a symmetric function on parameters $a,b,c,d$ and it suffices to
analyze only the case $a\geq b\geq c\geq d$. Obviously
$(ac-bd)\geq0,\,(ab-cd)\geq0$ and it remains  the constraint
$bc\geq ad$. The last inequality states that the product of the
largest and smallest coefficients must not exceed the product of
remaining coefficients. Denote by $s$ the smallest coefficient

\begin{equation}\label{ad.small}
s=\min(a,b,c,d).
\end{equation}

We can summarize all cases as follows

\begin{equation}\label{ad.largesmal}
l^2s^2\leq abcd.
\end{equation}

This is necessary but not sufficient condition for the existence
of $R_\times$. The next condition $S_\times^2>0$ we do not analyze
because the first condition (\ref{ad.largesmal}) suffices to
separate the validity domains.

\subsection{Separation of validity domains.}

In this section we define applicable domains of expressions
(\ref{cq.eigfin}),(\ref{cr.eigfin}) and (\ref{lc.larg}) step by
step.

\smallskip

\paragraph{Circumradius of convex quadrangle.} First we separate
the validity domains between the convex quadrangle and the largest
coefficient. In a highly entangled region, where the center of
circumcircle lies inside the quadrangle, the circumradius is
greater than any of sides and yield a correct answer. This situation
is changed when the center lies on the largest side of the
quadrangle and both equations (\ref{cq.eigfin}) and
(\ref{lc.larg}) give equal answers. Suppose that the side $a$ is
the largest one and the center lies on the side $a$. A little
geometrical speculation yields

\begin{equation}\label{ad.con-lar-a}
a^2=b^2+c^2+d^2+2\frac{bcd}{a}.
\end{equation}

From this equation we deduce that if $a^2$ is smaller than r.h.s., {\it i.e.}

\begin{equation}\label{ad.dom-lar-a}
a^2\leq b^2+c^2+d^2+2\frac{bcd}{a},
\end{equation}

\noindent then the circumradius-formula is valid. If $a^2$ is
greater than r.h.s in Eq.(\ref{ad.con-lar-a}), then the largest
coefficient formula is valid. The inequality (\ref{ad.dom-lar-a})
also guarantees the existence of the cyclic quadrilateral. Indeed,
using the inequality

\begin{equation}\label{ad.inter}
bc+cd+bd\geq3\frac{bcd}{a},
\end{equation}

\noindent one derives

\begin{equation}\label{ad.cycl}
(b+c+d)^2\geq b^2+c^2+d^2+\frac{6bcd}{a}\geq a^2.
\end{equation}

\noindent  Above inequality ensures the existence of a convex
quadrangle with the given sides.

To get a confidence, we can solve equation $u_k=\pm1$ using the
relation (\ref{ad.con-lar-a}). However, it is more transparent to
factorize it as following:

\begin{subequations}\label{ad.board1}
\begin{equation}\label{ad.board1+}
(4\omega^2-r_3^2)(1+u_k)=\frac{2ad}{bc+ad}
\left(b^2+c^2+d^2+\frac{2bcd}{a}-a^2\right)
\left(a^2+b^2+c^2+\frac{2abc}{d}-d^2\right)
\end{equation}
\begin{equation}\label{ad.board1-}
(4\omega^2-r_3^2)(1-u_k)=\frac{2bc}{bc+ad}
\left(a^2+c^2+d^2+\frac{2acd}{b}-b^2\right)
\left(a^2+b^2+d^2+\frac{2abd}{c}-c^2\right).
\end{equation}
\end{subequations}

Similarly, we have

\begin{subequations}\label{ad.board2}
\begin{equation}\label{ad.board2+}
(4\omega^2-r_3^2)(1+v_k)=\frac{2bd}{ac+bd}
\left(a^2+c^2+d^2+\frac{2acd}{b}-b^2\right)
\left(a^2+b^2+c^2+\frac{2abc}{d}-d^2\right)
\end{equation}
\begin{equation}\label{ad.board2-}
(4\omega^2-r_3^2)(1-v_k)=\frac{2ac}{ac+bd}
\left(b^2+c^2+d^2+\frac{2bcd}{a}-a^2\right)
\left(a^2+b^2+d^2+\frac{2abd}{c}-c^2\right).
\end{equation}
\end{subequations}

Thus, the circumradius of the convex quadrangle gives a correct
answer if all brackets in the above equations are positive.  In
general, Eq.(\ref{cq.eigfin}) is valid if

\begin{equation}\label{ad.con-lar}
l^2\leq\frac{1}{2}+\frac{abcd}{l^2}.
\end{equation}

When one of parameters vanishes, i.e. $abcd=0$, inequality
(\ref{ad.con-lar}) coincides with the corresponding condition in
Ref.\cite{lev}.

\smallskip

\paragraph{Circumradius of crossed quadrangle.} Next we
separate the validity domains between the convex and the crossed
quadrangles. If $S_\times^2<0$, then crossed one has no
circumcircle and the only choice is the circumradius of the convex
quadrangle. If $S_\times^2>0$, then we use the equality

\begin{equation}\label{ad.dif}
4R^2_q-4R^2_\times=\frac{r}{2}\frac{abcd}{S^2_qS^2_\times}
\end{equation}

\noindent where $r=r_1r_2r_3$. It shows that $r>0$ yields
$R_q>R_\times$ and vice-versa. Entanglement eigenvalue always
takes the maximal value. Therefore, $\lm=4R_q^2$ if $r>0$ and
$\lm=4R_\times^2$ if $r<0$. Thus $r=0$ is the separating surface
and it is necessary to analyze the condition $r<0$.

Suppose $a\geq b\geq c\geq d$. Then $r_2$ and $r_3$ are positive.
Therefore $r$ is negative if and only if $r_1$ is negative, which implies

\begin{equation}\label{ad.aplusd}
a^2+d^2>b^2+c^2.
\end{equation}

Now suppose $a\geq d\geq b\geq c$. Then $r_1$ is negative    and
$r_3$ is positive. Therefore $r_2$ must be positive, which implies

\begin{equation}\label{ad.aplusc}
a^2+c^2>b^2+d^2.
\end{equation}

It is easy to see that in both cases left hand sides contain
the largest and smallest coefficients. This result can be generalized
as follows: $r\leq0$ if and only if

\begin{equation}\label{ad.geq}
l^2\geq \frac{1}{2}-s^2.
\end{equation}

It remains to separate the validity domains between the
crossed-quadrangle and the largest coefficient. We can use three
equivalent ways to make this separation:

\smallskip

1)to use the geometric picture and to see when $4R_\times^2$ and
$l^2$ coincide,

2)directly factorize equation $u_k=\pm1$,

3)change the sign of the parameter $d$.

\smallskip

All of these give the same result stating that Eq.(\ref{cr.eigfin}) is
valid if

\begin{equation}\label{ad.leq}
l^2\leq\frac{1}{2}-\frac{abcd}{l^2}.
\end{equation}

Inequalities (\ref{ad.geq}) and (\ref{ad.leq}) together yield

\begin{equation}\label{ad.togeth}
l^2s^2\geq abcd.
\end{equation}

This inequality is contradicted by (\ref{ad.largesmal}) unless
$l^2s^2=abcd$. Special cases like $l^2s^2=abcd$ are considered
in the next section. Now we would like to comment the fact that
crossed quadrangle survives only in exceptional cases. Actually
crossed case can be obtained from the convex cases by changing the
sign of any parameter. It crucially depends on signs of parameters
or, in general, on phases of parameters. On the other hand all
phases in Eq.(\ref{w.psi}) can be eliminated by
LU-transformations. For example, the phase of $d$ can be
eliminated by redefinition of the phase of the state function
$|\psi\ra$ and the phases of remaining parameters can be absorbed
in the definitions of basis vectors $|1\ra$ of the qubits A, B and
C. Owing to this entanglement eigenvalue being LU invariant
quantity does not depend on phases. However, crossed case is
relevant if one considers states given by Generalized Schmidt
Decomposition(GSD) \cite{acin}. In this case phases can not be
gauged away and crossed case has its own range of definition. This
range has shrunk to the separating surface $r=0$ in our case.

\medskip

Now we are ready to present a distinct separation of the validity
domains:

\begin{equation}\label{ad.full}
\lm=
\begin{cases}
 \enskip 4R_q^2, &{\rm if}\quad l^2\leq1/2+abcd/l^2\cr
 \enskip l^2 &{\rm if}\quad l^2\geq1/2+abcd/l^2
\end{cases}
\end{equation}

As an illustration we present the plot of $d$-dependence of $\lm$ in
Fig.2 when $a=b=c$.

\begin{figure}
\includegraphics[width=10cm]{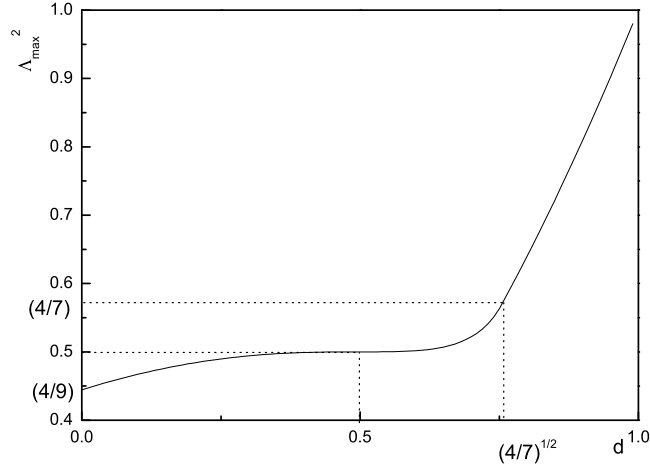}
\caption{\label{sim.eps}Plot of $d$-dependence of
$\Lambda_{max}^2$ when $a=b=c$. When $d\to1$, $\lm $ goes to $1$
as expected. When $d=0$, $\lm$ becomes $4/9$, which coincides with
the result of Ref.\cite{wei}. When $r=0$ which implies $a = d =
1/2$, $\lm$ becomes $1/2$ (it is shown as dotted line). When $d =
2 a$, which implies $d = \sqrt{4/7}$, $\lm$ goes to $4/7$, which
is one of shared states (it is also shown as another dotted
line).}
\end{figure}

We have distinguished three types of quantum states depending on
which expression takes entanglement eigenvalue. Also there are
states that lie on surfaces separating different applicable
domains. They are shared by two types of quantum states and may
have interesting features. We will call those shared states. Such
shared states are considered in the next section.

\section{Shared States.}

Consider quantum states for which both convex and crossed
quadrangles yield the same entanglement eigenvalue.
Eq.(\ref{cr.ab}) is not applicable and we rewrite equations
(\ref{cq.radius}) and (\ref{cr.radius}) as follows

\begin{equation}\label{ss.con-cr}
4R_q^2=\frac{1}{2}\left(1-\frac{r}{16S^2_q}\right),\quad
4R_\times^2=\frac{1}{2}\left(1-\frac{r}{16S^2_\times}\right).
\end{equation}

These equations show that if the state lies on the separating
surface $r=0$, then entanglement eigenvalue is a constant

\begin{equation}\label{ss.shar1}
\lm=\frac{1}{2}
\end{equation}

\noindent and does not depend on the state parameters. This fact
has a simple interpretation. Consider the case $r_1=0$. Then
$b^2+c^2=a^2+d^2=1/2$ and the quadrangle consists of two right
triangles. These two triangles have a common hypotenuse and legs
$b,c$ and $a,d$, respectively, regardless of the triangles being
in the same semicircle or in opposite semicircles. In both cases
they yield same circumradius. Decisive factor is that the center
of the circumcircle lies on the diagonal. Thus the perimeter and
diagonals of the quadrangle divide ranges of definition of the
convex quadrangle. When the center of circumcircle passes the
perimeter, entanglement eigenvalue changes-over from convex
circumradius to the largest coefficient. And if the center lies on
the diagonal, convex and crossed circumradiuses become equal.

We would like to bring plausible arguments that this picture is
incomplete and there is a region that has been shrunk to the point.
Consider three-qubit state given by GSD

\begin{equation}\label{ss.gsd}
|\psi\ra=a|100\ra+b|010\ra+b|001\ra+d|111\ra+e|000\ra.
\end{equation}

One of parameters must have non-vanishing phase\cite{acin} and we
can treat this phase as an angle. Then, we have five sides and an
angle. This set defines a  sexangle that has circumcircle. One can
guess that in a highly entangled region entanglement eigenvalue is
the circumradius of the sexangle. However, there is a crucial
difference. Any convex sexangle contains a star type area and the
sides of this area are the diagonals of the sexangle. The
perimeter of the star separates the convex and the crossed cases.
Unfortunately, we can not see this picture in our case because the
diagonals of a quadrangle confine a single point. It is left for
future to calculate the entanglement eigenvalues for arbitrary
three qubit states and justify this general picture.

Shared states given by $r=0$ acquire new properties. They can be
used for perfect teleportation and superdense coding
\cite{lev,agr,tele}. This statement is not proven clearly, but also
no exceptions are known.

\medskip

Now consider a case where the largest coefficient and circumradius of
the convex quadrangle coincide with each other. The separating
surface is given by

\begin{equation}\label{ss.con-lar}
l^2=\frac{1}{2}+\frac{abcd}{l^2}.
\end{equation}

Entanglement eigenvalue ranges within the narrow interval

\begin{equation}\label{ss.shar2}
\frac{1}{2}\leq\lm\leq\frac{4}{7}.
\end{equation}

It separates slightly and highly entangled states. When one of
coefficients is large enough and satisfies the relation
$l^2>1/2+abcd/l^2$, entanglement eigenvalue takes a larger
coefficient. And the expression (\ref{w.psi}) for the state
function effectively takes the place of Schmidt decomposition. In
highly entangled region no similar picture exists and all
coefficients participate in equal parts and yield the circumradius.
Thus, shared states given by Eq.(\ref{ss.con-lar}) separate
slightly entangled states from highly entangled ones, and
can be ascribed to both types.

What is the meaning of these states? Shared states given by $r=0$
acquire new and important features. One can expect that shared
states dividing highly and slightly entangled states also must
acquire some new features.
However, these features are yet to be discovered.

\section{Conclusions}

We have considered three-parametric families of three qubit states
and derived explicit expressions for entanglement eigenvalue. The
final expressions have their own geometrical interpretation. The
result in this paper with the results of Ref.\cite{lev}
show that the geometric measure has two visiting cards:
the circumradius and the largest coefficient. The geometric interpretation
may enable us to predict the answer for the states given by GSD.
If the center of circumcircle lies in star type area confined by
diagonals of the sexangle, then entanglement eigenvalue is the
circumradius of the crossed sexangle(s). If the center lies in the
remaining part of sexangle, the entanglement eigenvalue is the
circumradius of the convex sexangle. And when the center passes
the perimeter, then entanglement eigenvalue is the largest
coefficient. Although we cannot justify our prediction
due to lack of computational technique, this picture surely
enables us to take a step toward a deeper
understanding of the entanglement measure \cite{3qub}.

Shared states given by $r=0$ play an important role in quantum
information theory. The application of shared states given by
Eq.(\ref{ss.con-lar}) is somewhat questionable, and should be
analyzed further. It should be pointed out that one has to
understand the properties of these states and find the possible
applications. We would like to investigate this issue elsewhere.

Finally following our procedure, one can obtain the nearest
product state of a given three-parametric W-type state. These two
states will always be separated by a line of densities composed of
the convex combination of W-type states and the nearest product
states \cite{pit}. There is a separable density matrix $\varrho_0$
which splits the line into two parts as follows. One part consists
of separable densities and another part consists of  non-separable
densities. It was shown in Ref.\cite{pit} that an operator
$W=\varrho_0-\rho^{ABC}-\tr[\varrho_0(\varrho_0-\rho^{ABC})]I$ has
the properties $\tr(W\rho^{ABC})<0$, and $\tr(W\varrho)\geq0$ for
the arbitrary separable state $\varrho$. The operator $W$ is
clearly Hermitian and thus is an entanglement witness for the
state. Thus our results allow oneself to construct the
entanglement witnesses for W-type three qubit states. However, the
explicit derivation of $\varrho_0$ seems to be highly non-trivial
\cite{sper,kram}.

\begin{acknowledgments}
LT thanks Edward Chubaryan for help. This work was supported by
the Kyungnam University Research Fund, 2007.
\end{acknowledgments}

\end{document}